\documentclass[aps,prd,preprint,superscriptaddress,tightenlines,nofootinbib]{revtex4}

\usepackage{graphicx}
\usepackage{dcolumn}
\usepackage{bm}
\usepackage{amsmath}

\newcommand{\ra}{\ensuremath{\rightarrow}}

\newcommand{\GeVUnit}{\ensuremath{{\rm GeV}}}
\newcommand{\MomUnit}{\ensuremath{{\rm GeV}/c}}

\newcommand{\MassUnit}{\ensuremath{{\rm GeV}/c^2}}
\newcommand{\MassSqrUnit}{\ensuremath{{\rm GeV}^2/c^4}}
\newcommand{\MassFourthUnit}{\ensuremath{{\rm GeV}^4/c^8}}

\newcommand{\Vub}{\ensuremath{|V_{ub}|}}
\newcommand{\Vcb}{\ensuremath{|V_{cb}|}}

\newcommand{\Blnu}[1]{\ensuremath{B \ra {#1} l \nu}}
\newcommand{\BXlnu}{\Blnu{X}}
\newcommand{\BXulnu}{\Blnu{X_u}}
\newcommand{\BXclnu}{\Blnu{X_c}}

\newcommand{\BDlnu}{\Blnu{D}}
\newcommand{\BDSlnu}{\Blnu{D^*}}
\newcommand{\BDSSlnu}{\Blnu{D^{**}}}

\newcommand{\bsg}{\ensuremath{B \ra X_s \gamma\ }}

\newcommand{\Elep}{\ensuremath{E_{\ell}}}

\newcommand{\Enu}{\ensuremath{E_{\nu}}}

\newcommand{\QSqr}{\ensuremath{q^2}}

\newcommand{\MX}{\ensuremath{M_{X}}}
\newcommand{\MXSqr}{\ensuremath{M_{X}^2}}
\newcommand{\CosWL}{\ensuremath{\cos\theta_{W\ell}}}

\newcommand{\Bbd}{\ensuremath{{\hbox{$\overline{B^0}$}\kern-1.45em\raise 1.7ex
          \hbox{\tiny \bf ( \hskip1.3em )}}}}

\newcommand{\nubar}{\ensuremath{\overline{\nu}}}
\newcommand{\BBbar}{\ensuremath{B \overline B}}
\newcommand{\qqbar}{\ensuremath{q \overline q}}
\newcommand{\qbar}{\ensuremath{\overline q}}
\newcommand{\BRsym}{{\cal B}}

\newcommand{\MomOf}[1]{\ensuremath{\langle#1\rangle}}
\newcommand{\MDbarSqr}{\ensuremath{{\overline M}_D^2}}
\newcommand{\MomMXSqrmMDbSqr}{\MomOf{\MXSqr-\MDbarSqr}}

\newcommand{\Lambar}{\ensuremath{\overline\Lambda}}
\newcommand{\lam}[1]{\ensuremath{\lambda_{#1}}}

\def\Journal#1&#2&#3(#4){ {\it #1}{\bf #2}, #3 (#4)}

\newcommand{\mypsf}[3]{
\begin{figure}[hptb]
\begin{center}
  #2
\caption{#3}
\label{#1}
\end{center}
\end{figure}
}

\begin{document}

\preprint{CLEO-CONF 03-08}
\preprint{LP-279}

\title{Measurement of the Hadronic Recoil Mass Moments in Semileptonic $B$ Decay}

\thanks{Submitted to XXI International Symposium on Lepton and Photon Interactions at High Energies, August 11-16, 2003, Fermi National Accelerator Laboratory, Batavia, IL.}


\author{G.~S.~Huang}
\author{D.~H.~Miller}
\author{V.~Pavlunin}
\author{B.~Sanghi}
\author{E.~I.~Shibata}
\author{I.~P.~J.~Shipsey}
\affiliation{Purdue University, West Lafayette, Indiana 47907}
\author{G.~S.~Adams}
\author{M.~Chasse}
\author{J.~P.~Cummings}
\author{I.~Danko}
\author{J.~Napolitano}
\affiliation{Rensselaer Polytechnic Institute, Troy, New York 12180}
% ...
\collaboration{CLEO Collaboration} 
\noaffiliation


\date{\today}

\begin{abstract} 

     We present a preliminary measurement of the composition of inclusive semileptonic $B$ 
     meson decays
     using $9.4\ {\rm fb}^{-1}$ of $e^+e^-$ data taken with the CLEO detector at the $\Upsilon(4S)$ 
     resonance. In addition to measuring the charged lepton kinematics, the neutrino 4-vector is 
     inferred using the hermiticity of the detector. We perform a maximum likelihood fit
     over the full three-dimensional differential decay distribution for
     the fractional contributions from the $B\rightarrow X_cl \nu$ processes with $X_c = D$,
     $D^*$, $D^{**}$, and nonresonant $X_c$, and the process $B\rightarrow X_u l \nu$. 
     From the fit results we extract $\MomMXSqrmMDbSqr = (0.456 \pm 0.014 \pm 0.045 \pm 0.109)\ \MassSqrUnit$
     with minimum lepton energy of \mbox{1.0 \GeVUnit}\ and $\MomMXSqrmMDbSqr = 
     (0.293 \pm 0.012 \pm 0.033 \pm 0.048)\ \MassSqrUnit$
     with minimum lepton energy of \mbox{1.5 \GeVUnit}. The uncertainties are from statistics, detector
     systematic effects, and model dependence, respectively.
    
\end{abstract}

\maketitle


\section{Introduction}

Nonperturbative QCD physics connects measurements of $B$ meson
decay properties to the properties of underlying weak 
flavor changing currents. Heavy-quark effective theory (HQET) combined
with the operator product expansion (OPE) provides a framework
in which many inclusive $B$ decay properties can be calculated 
\cite{ref:general_HQET_OPE}.

Using HQET, the inclusive $B$ decay matrix elements can be expanded 
in powers of $\Lambda_{QCD}/M_B$. For each order in the 
expansion new nonperturbative parameters arise. At order $\Lambda_{QCD}/M_B$,
there is \Lambar and at order $\Lambda_{QCD}^2/M_B^2$, there
are \lam1 and \lam2. The nonperturbative parameter \Lambar\ relates the 
b-quark mass to the $B$ meson mass in the limit of infinite b-quark mass. 
 The parameter \lam2
is directly related to the mass splitting between the $B^*$ and $B$ mesons.
Moments of kinematic variables can be used to determine
the \Lambar\ and \lam1\ parameters.
These parameters are properties of the B meson and are not specific to the
decay mode being studied. This means that moments of the lepton energy and
the $X_c$ mass in \BXclnu\ decays can be related to moments of the photon 
energy spectrum in \bsg.  Furthermore, measurements of these parameters
can be used in the calculation of absolute decay rates, providing an 
improved relationship between the \BXclnu\ branching fraction and \Vcb.

This paper presents a measurement of the first moment of the square of the
hadronic mass in \BXclnu\ decays as a function of the lepton energy cut. Because the
background from secondary leptons becomes large at low lepton energy,
the lowest lepton energy considered is 1.0 \GeVUnit. 
Moments of the square of the hadronic mass have previously been measured by CLEO with
a lepton energy cut of 1.5 \GeVUnit\ \cite{ref:oldhadmom} and by BaBar
with a lepton energy cut of 0.9 \GeVUnit\ \cite{ref:babarmoms}.
As the lepton energy cut is made more restrictive, the terms that are higher order in 
$\Lambda_{QCD}/M_B$ become more important,
making the expansion less reliable \cite{ref:falkcut}. There are significant complications in 
reducing the minimum lepton energy to 1.0 \GeVUnit. The backgrounds from 
continuum $e^+e^-\ra q\qbar$ events, fake leptons and secondaries all become more important. 
The structure of the nonresonant \BXclnu\ decays also becomes more important.

In order to reconstruct the recoiling hadronic invariant mass, it is 
necessary to reconstruct both the charged lepton and neutrino
kinematics. The neutrino is reconstructed using the
approximate hermiticity of the CLEO II  and CLEO II.V detectors
\cite{ref:NIM} and the well known initial state of the $e^+e^-$ system 
produced by the Cornell Electron Storage Ring (CESR) \cite{ref:firstnurec}.
 The experimental resolution
on the neutrino energy has a narrow core with a full width at half maximum height
of approximately 120 MeV and a broad
tail of over estimation of the neutrino energy which extends up to 1.5 \GeVUnit. The
square of the hadronic recoil mass, \MXSqr, can be calculated if the $B$ momentum, 
$\vec{p}_B$, is known,
\begin{eqnarray*}
\label{mxeqn}
\MX^2 &=& M_B^2 + \QSqr - 2 E_{beam}(E_{\ell}+E_{\nu}) +
 2 |\vec{p}_B| |\vec{p}_{\ell}+\vec{p}_{\nu}| \cos\theta_{B\cdot \ell\nu},
\end{eqnarray*}
\noindent
where $\theta_{B\cdot \ell\nu }$ is the angle between the momentum of
the leptonic system, $\vec{q}=\vec{p}_{\ell}+\vec{p}_{\nu}$, and $\vec{p}_B$. 
Since the $B$ mesons are the daughters of an $\Upsilon(4S)$ 
produced at rest, the magnitude of the $B$ momentum is known and small, however
its direction is unmeasured. The last term in the \MXSqr\ equation depends on the
unknown $B$ momentum direction, but it is small, and therefore neglected in this analysis.  
Because of the neglected term, the \MXSqr\ resolution depends on $|\vec{q}|$. 
The kinematic variables $\QSqr=(p_{\ell}+p_{\nu})^2$ and the pseudo-helicity angle
of the virtual W, \CosWL, can also be calculated. The pseudo-helicity angle, \CosWL,
is defined as the cosine of the angle between the lepton momentum in the virtual-W
frame and the virtual-W momentum in the lab frame.

From the sample of events with a charged lepton
and reconstructed neutrino, the differential decay rate
is measured as a function of the reconstructed quantities \QSqr, \MXSqr, and \CosWL.
We fit the observed three-dimensional distribution to a sum of Monte Carlo models for the 
different hadronic final states: $D$, $D^*$, $D^{**}$, $X_c$ nonresonant, and $X_u$.
The models are constructed with Monte Carlo events generated with a full detector
simulation \cite{ref:GEANT} and have the same reconstruction cuts applied as the data.
In addition to the \BXlnu\ processes, there are backgrounds from fake leptons,
continuum leptons and secondary leptons. The \QSqr\ and \CosWL\ variables add
information which contributes to the separation of the signal from the backgrounds,
and the separation of the $D$ and $D^*$ final states.
The moment \MomMXSqrmMDbSqr\ is calculated from the fit results for the 
final states containing charm using the same theoretical models as are used
in the fit, but without any detector simulation.

\section{Data Sample and Event Selection}

The data used in this analysis were taken with two configurations of the 
CLEO detector, CLEO II and CLEO II.V. An integrated luminosity of
$9.4\ {\rm fb}^{-1}$ was accumulated on the $\Upsilon(4S)$ resonance, $E_{cm}\approx 10.58\ \GeVUnit$ 
and an additional $4.5\ {\rm fb}^{-1}$ was accumulated at 60 MeV below the $\Upsilon(4S)$ resonance,
where there is no \BBbar\ production. Both detector configurations covered 95\% of 
the 4$\pi$ solid angle with drift chambers and a cesium iodide
calorimeter. Particle identification was provided by muon chambers with measurements 
made at 3, 5, and 7 hadronic interaction lengths, a time of flight system and
specific ionization ($dE/dx$) from the drift chamber. In the CLEO II configuration, 
there were three concentric drift chambers filled with a mixture of argon and ethane.
In the CLEO II.V detector, the innermost tracking chamber was replaced with a 
three-layer silicon detector and the main drift chamber gas was changed to a mixture of 
helium and propane.

Events are selected to have an identified electron or muon with momentum
greater than 1 \MomUnit\ and a well reconstructed neutrino. Additional criteria are used
to suppress background events from the $e^+e^- \ra q\qbar$ continuum under the
$\Upsilon(4S)$ resonance.

The identified leptons are required to fall within the barrel region of the detector 
($|\cos\theta|<0.71$, where $\theta$ is the angle between the lepton momentum 
and the beam axis).
Electrons are identified with a likelihood-based discriminator which combines
$dE/dx$, time of flight, and the ratio of the energy deposited in the calorimeter
to the momentum of the associated charged track ($E/p$). Muons are identified by their
penetration into the muon chambers. For momenta between 1.0 and 1.5 \MomUnit, muon 
candidates are required
to penetrate at least 3 interaction lengths and above 1.5 \MomUnit, candidates are
required to penetrate at least 5 interaction lengths. The
absolute lepton identification efficiencies are calculated by embedding raw data from 
reconstructed leptons in radiative QED events into hadronic events. The rate at which 
pions and kaons
fake leptons is measured by reconstructing $K^0_S \ra \pi^+\pi^-$, $D^0 \ra K^-\pi^+$,
and $\overline{D^0} \ra K^+\pi^-$ using only kinematics and then 
checking the daughter particle lepton identification information.


Neutrinos are reconstructed by subtracting the sum of the four-momenta of all observed 
tracks  and showers not associated with tracks, $p_{observed}^\mu$, from the 
four-momentum of the $e^+ e^-$  initial state, $p_{e^+ e^-}^\mu$, which is nearly at 
rest in the laboratory, 
\begin{eqnarray*}
p_\nu^\mu = p_{e^+ e^-}^\mu - p_{observed}^\mu.
\end{eqnarray*}
The errors made in this assumption are due to particles lost through inefficiency or
limited acceptance, fake tracks and showers, and other undetected particles such as
$K^0_L$ mesons, neutrons, or additional neutrinos. Several requirements are made to select
events in which these effects are reduced and the neutrino four-momentum resolution
is correspondingly enhanced. 

Because extra neutrinos are  correlated with extra leptons, events with an identified lepton
in addition to the signal lepton are rejected. The primary source of fake
tracks is from charged particles which do not have sufficient transverse momentum
to reach the calorimeter and therefore curl in the tracking chambers, returning
to the beam axis.
The portions of the track after the initial outbound portion may be accidentally 
included as additional tracks. Criteria have been developed to identify such errors and
make a best estimate of the actual charged particles in the event. Events for
which the total charge of the tracks is not zero are removed, reducing the 
effect of lost or fake tracks. 
Showers in the calorimeter associated with tracks in the drift chamber are not used,
so as not to double count their energy.  A neural network algorithm has been 
developed to reject secondary hadronic showers associated with showers that are 
already associated with tracks. 

A final neutrino reconstruction quality requirement is that the mass of the 
reconstructed neutrino to be small. The ratio of the reconstructed neutrino 
invariant mass squared, $M_{\rm miss}^2$, to twice the reconstructed neutrino energy 
is required to satisfy $|{{M_{\rm miss}^2}}/{2E_\nu}|<0.35\ \mbox{GeV}/c^4$. 
This quantity is proportional to the energy of a lost or fake 
particle. After this cut, the reconstructed neutrino's energy is assigned to be 
the magnitude of the missing momentum, because the momentum is not dependent on the particle 
identification of the tracks and so has a better resolution than the direct 
energy measurement.

Continuum events are suppressed by a combination of event shape and orientation
criteria which exploit the fact that continuum events tend to be jet-like with
a jet axis aligned with the beam axis, whereas \BBbar\ events are more spherical and
their orientation is uniformly distributed in the detector. 
The second Fox-Wolfram moment \cite{ref:foxwolfram}, $R_2$, of the energy flow in the event is
required to be less than 0.4. In addition, a neural network is used to combine $R_2$, 
the angle between the lepton and the event thrust axis, 
the angle between the lepton momentum and the beam axis, and the fraction of the total energy
lying in nine separate cones around the lepton direction, which cover the full the $4\pi$
solid angle. The $R_2$ cut is more than 99\% and 95\% efficient for \BXclnu\ and \BXulnu, respectively,
while removing 60\% of the continuum events. The neural net cut removes an additional
73\% of the continuum background, while keeping 92\% and 94\% of the \BXclnu\ 
and \BXulnu, respectively.

After all cuts we observe 41411 events from CLEO II and 80440 events from CLEO II.V.
The overall efficiency varies from 1.5\% for \BXclnu\ nonresonant to 4.2\% for \BXulnu.

\section{Fit Technique}

The full three-dimensional differential decay rate distribution as a function
of the reconstructed quantities \QSqr, \MXSqr, and \CosWL\ is fit for the 
contributions from semileptonic $B$ decay and backgrounds.
The \QSqr\ variable is replaced by $\QSqr / (E_\ell + E_\nu )^2$ for fitting 
purposes.
The $B$ decay modes are \BDlnu, \BDSlnu, \BDSSlnu, \BXclnu\  nonresonant, and \BXulnu.
The backgrounds are classified as secondary leptons, continuum leptons, or
fake leptons. A secondary lepton is a real lepton in a \BBbar\ event whose parent
is not a $B$ meson. A continuum lepton is a real lepton in a continuum 
$e^+e^- \ra \qqbar$ event. A fake lepton is a non-leptonic track from either a \BBbar\ 
or a continuum event which is identified as a lepton. 

The normalization of the continuum lepton component
is determined from the data taken below \BBbar\ threshold.
The normalization of the fake leptons is determined from the measured fake
rates and the measured track spectra. The contributions of these two backgrounds 
are therefore not allowed to vary in the fit, while those of the secondary leptons
and all of the \BXlnu\ modes are.

We perform a binned maximum-likelihood fit where component histograms 
are constructed from weighted Monte Carlo or data events.
The fit uses electrons and muons simultaneously, with a separate set 
of histograms for each. The likelihood is implemented to take into
account the histogram statistics using the method described in reference 
\cite{ref:BarlowBeeston}. Projections of the Monte Carlo simulations of
reconstructed quantities \QSqr, \MXSqr, and \CosWL\ for the various 
\BXlnu\ modes are shown in Figure \ref{fig:btox}. Projections of the data
and fit result are shown in Figure \ref{fig:fitprojs}.

\mypsf{fig:btox}
{
  \resizebox{.38\textwidth}{!}{\includegraphics{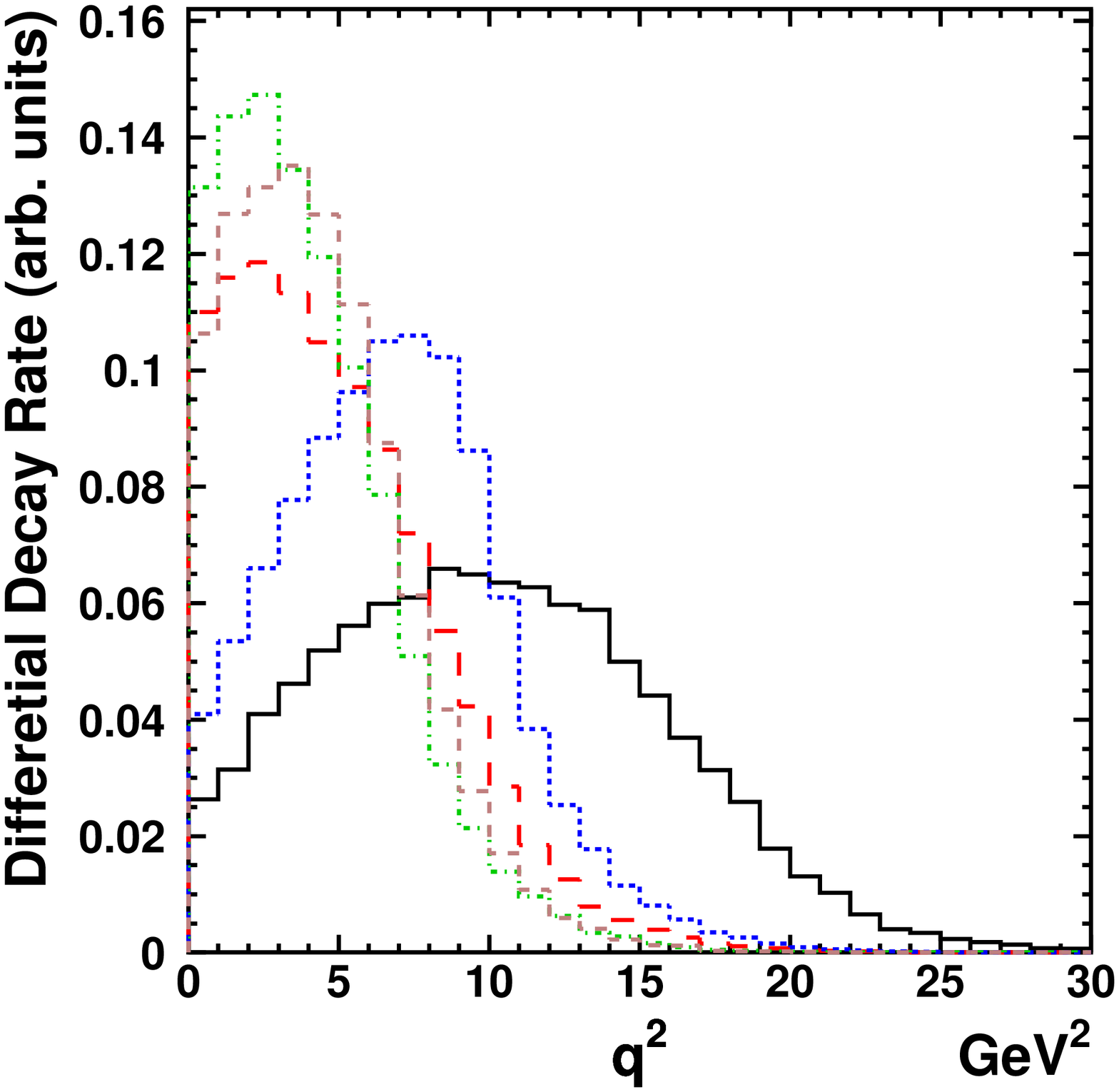}}
  \resizebox{.38\textwidth}{!}{\includegraphics{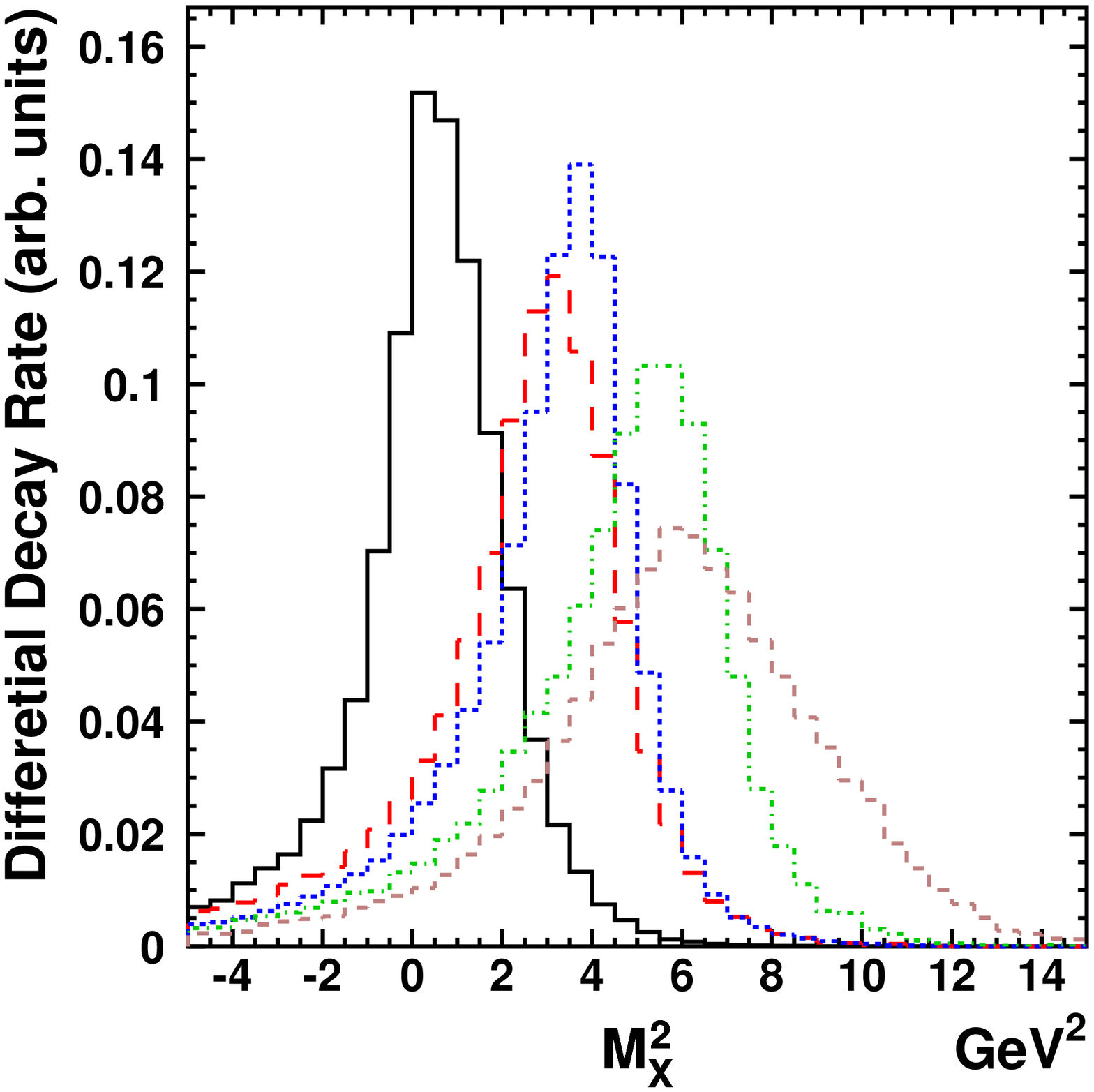}}
  \resizebox{.38\textwidth}{!}{\includegraphics{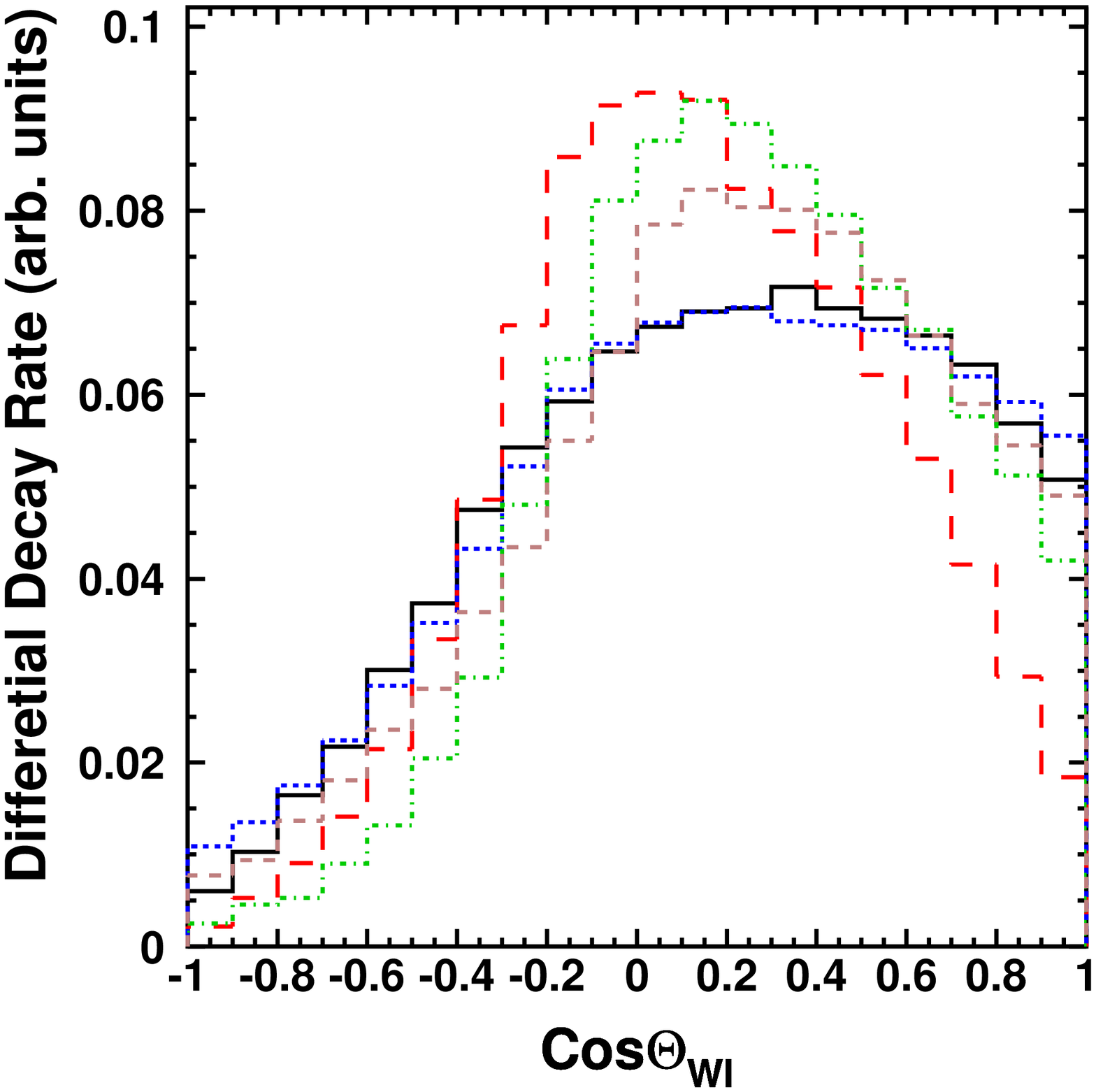}}
  \resizebox{.38\textwidth}{!}{\includegraphics{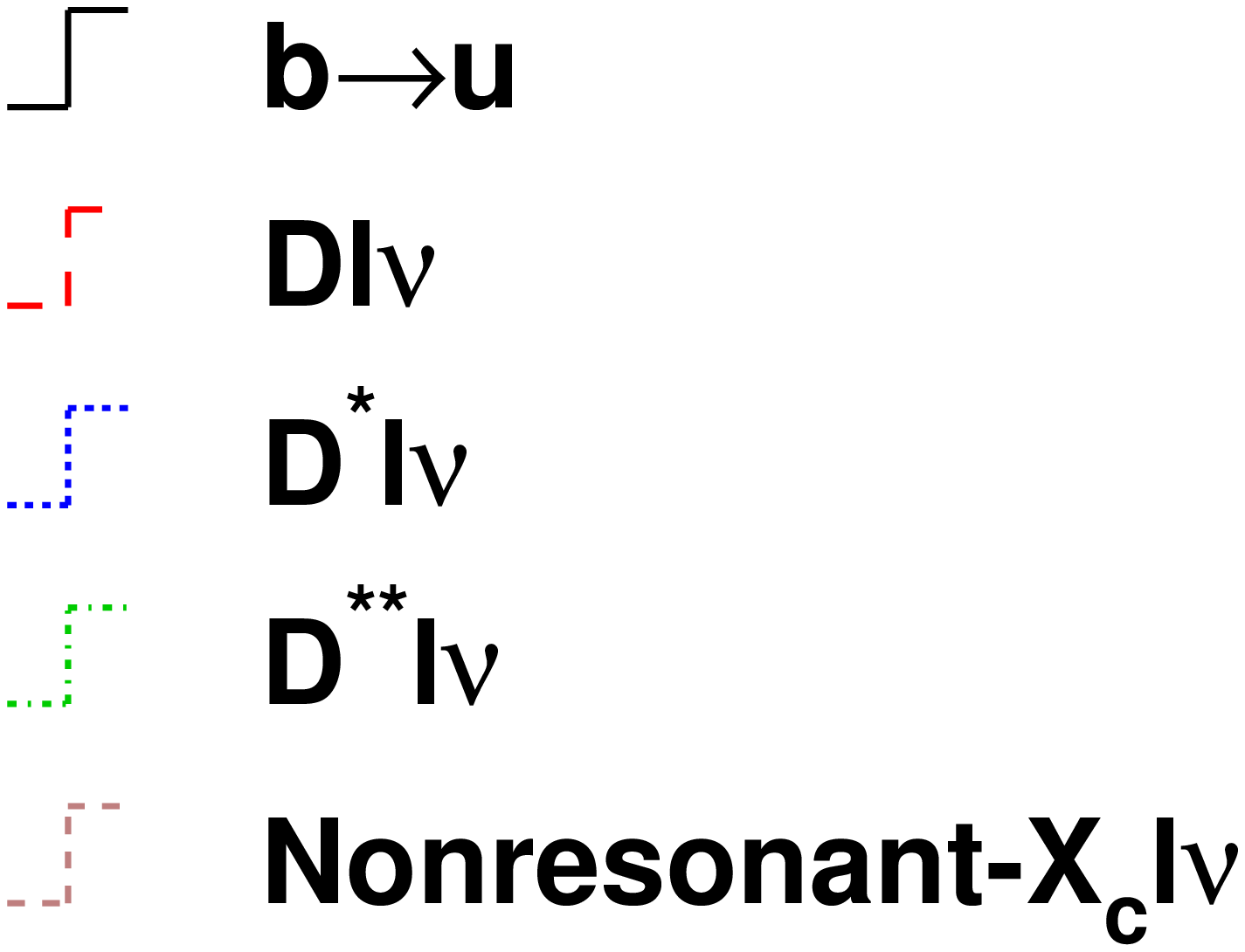}}
}
{
  Monte Carlo simulation of the distributions of the reconstructed quantities \QSqr, \MXSqr,
  and \CosWL\ for the various \BXlnu\ modes.
  The modes are \BXulnu\  (solid), \BDlnu\ (short dash), \BDSlnu\ (dots), \BDSSlnu\ (dot-dash),
  and \BXclnu\  nonresonant (long dash). The curves are
  normalized to have unit area to facilitate comparison of the shapes.
  Note that due to finite resolution \MXSqr\ can be less than zero. 
}

\mypsf{fig:fitprojs}
{
  \resizebox{.38\textwidth}{!}{\includegraphics{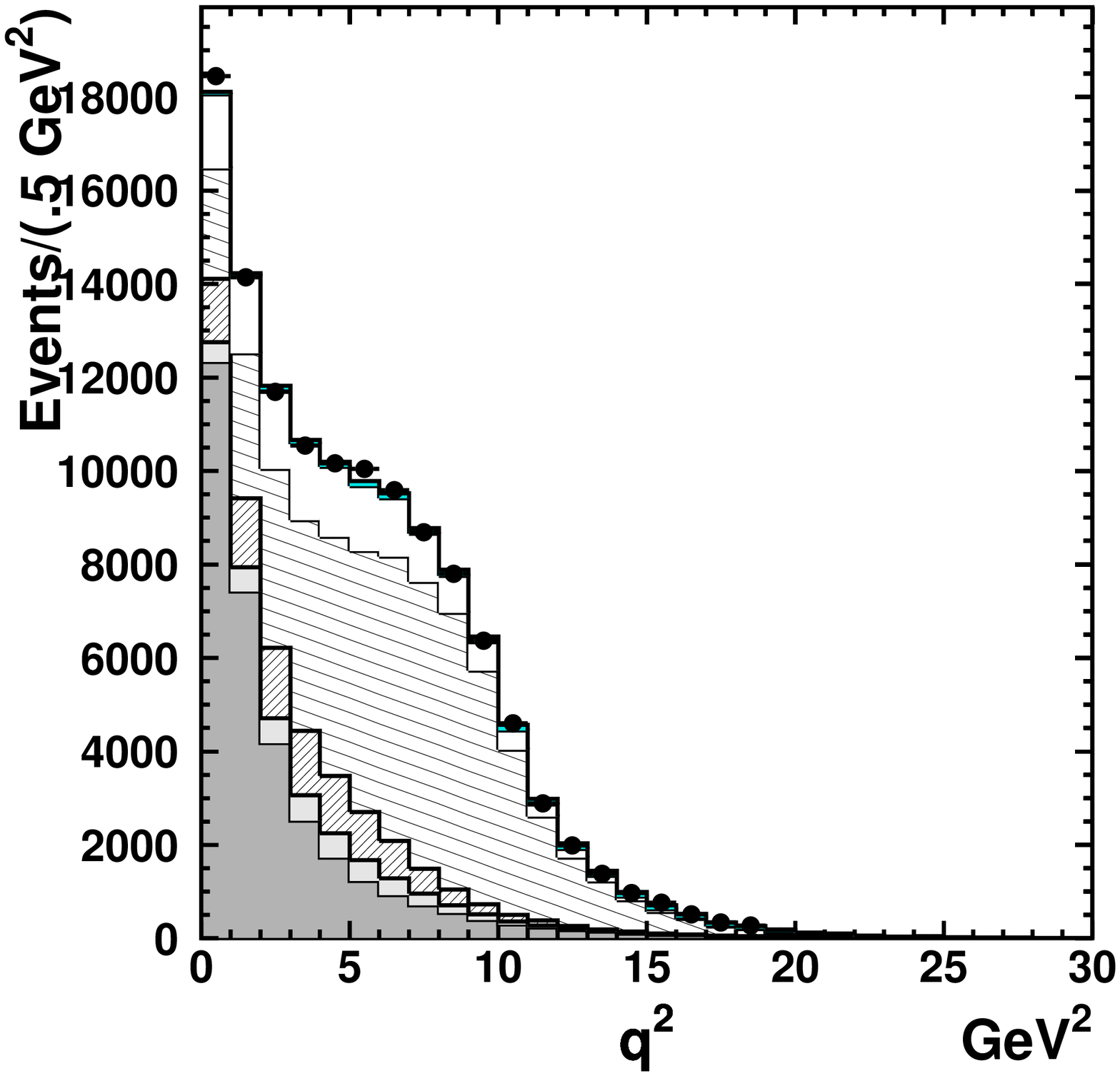}}
  \resizebox{.38\textwidth}{!}{\includegraphics{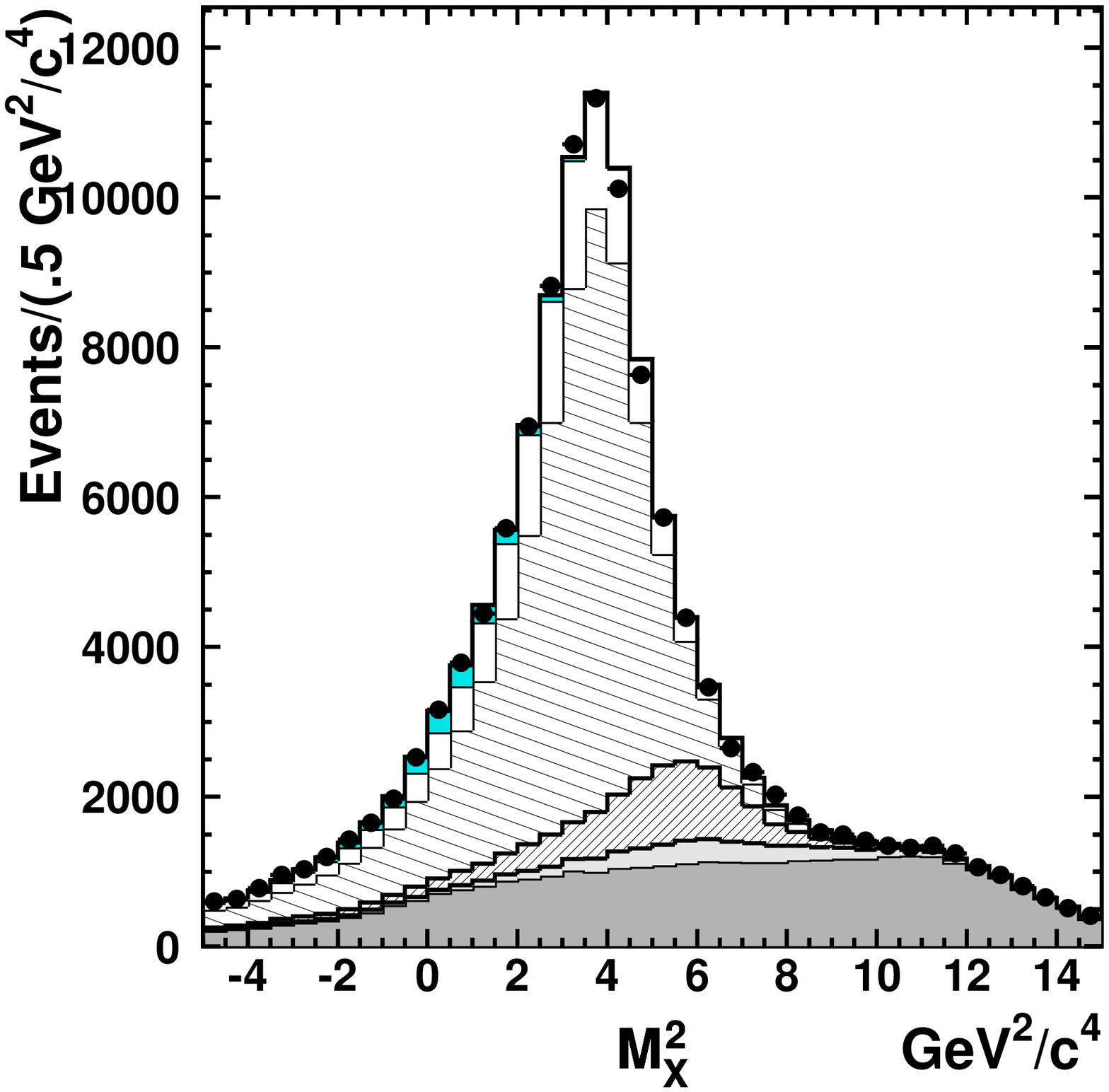}}
  \resizebox{.38\textwidth}{!}{\includegraphics{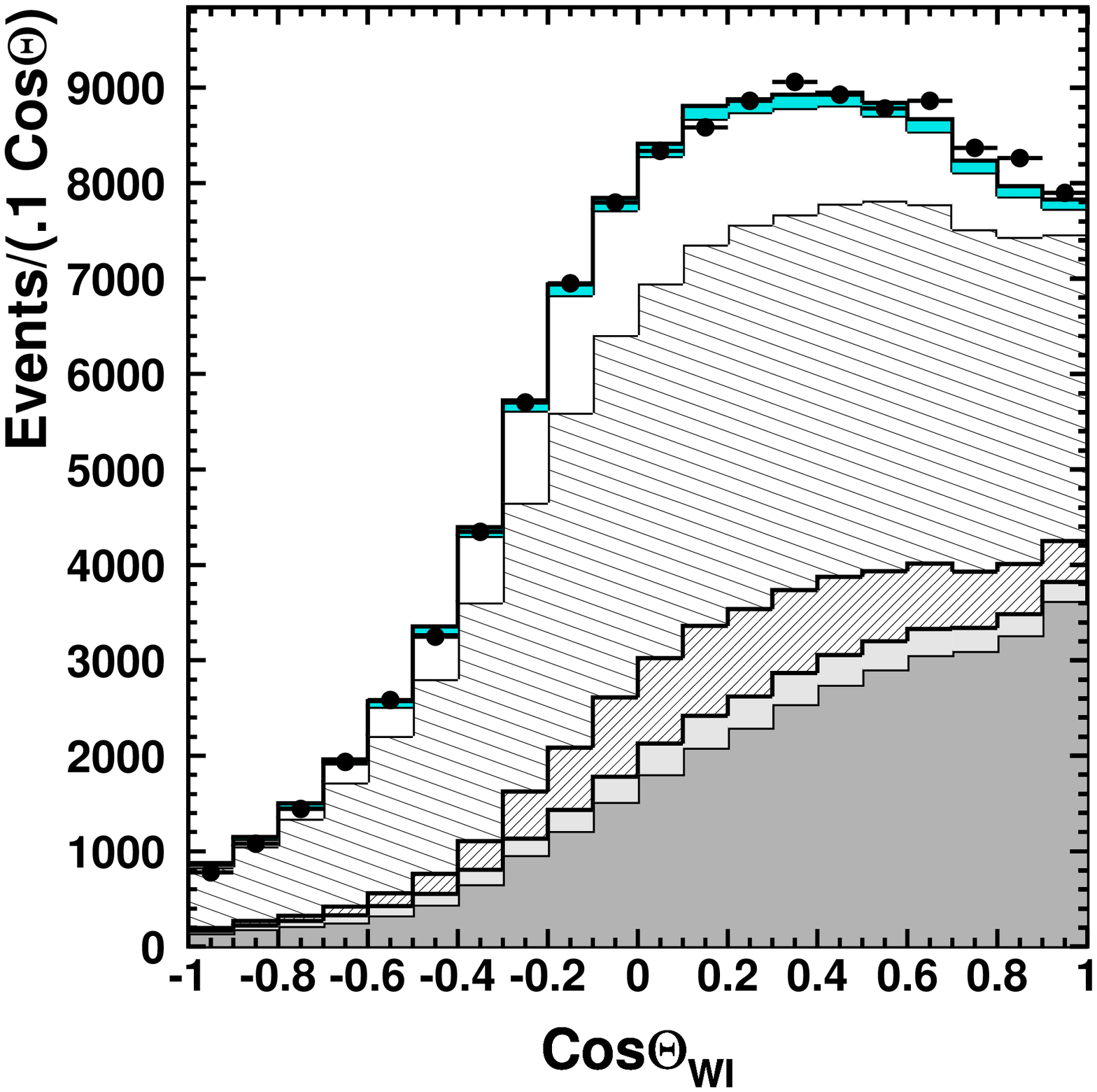}}
  \resizebox{.38\textwidth}{!}{\includegraphics{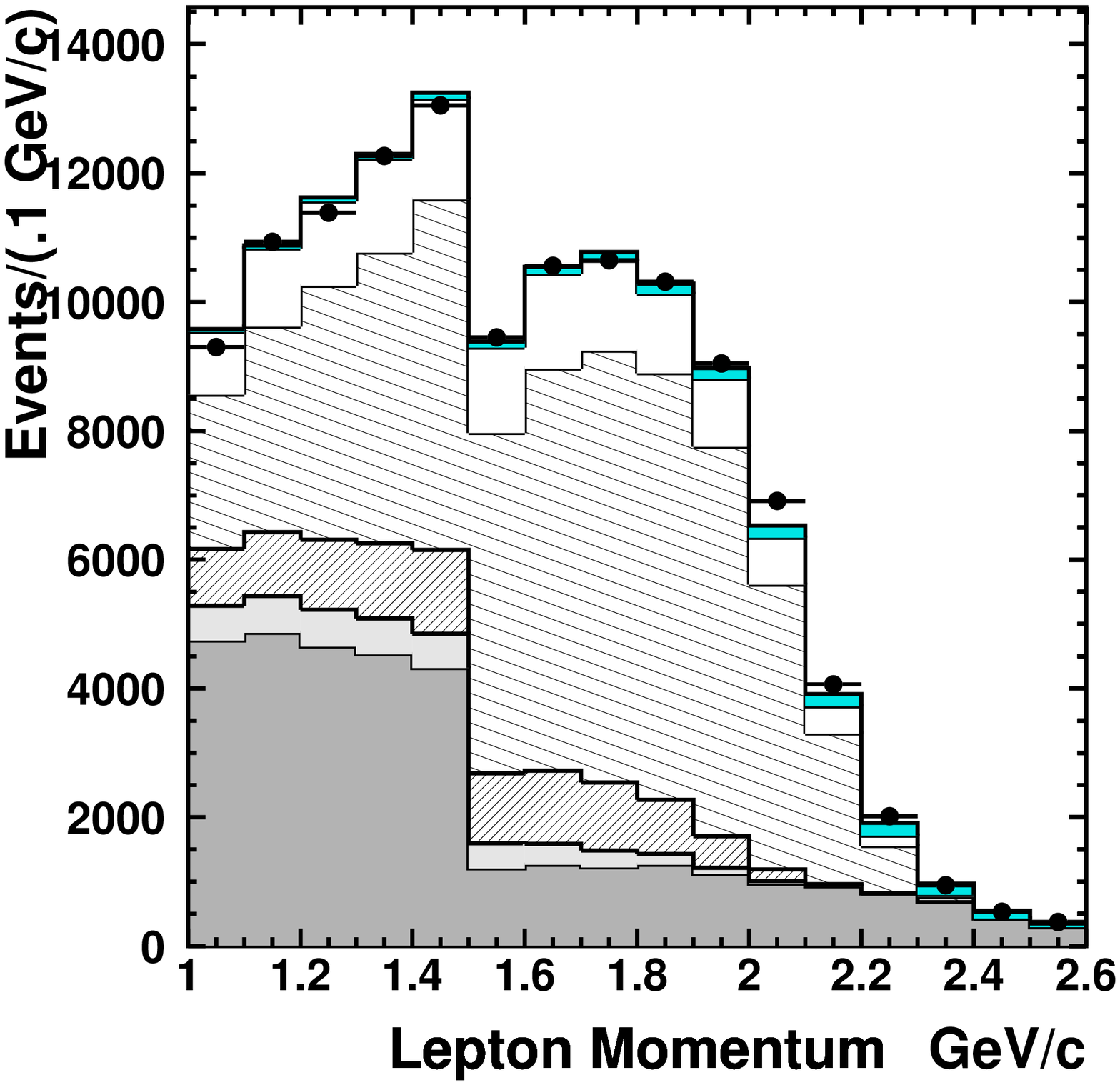}}
  \resizebox{.38\textwidth}{!}{\includegraphics{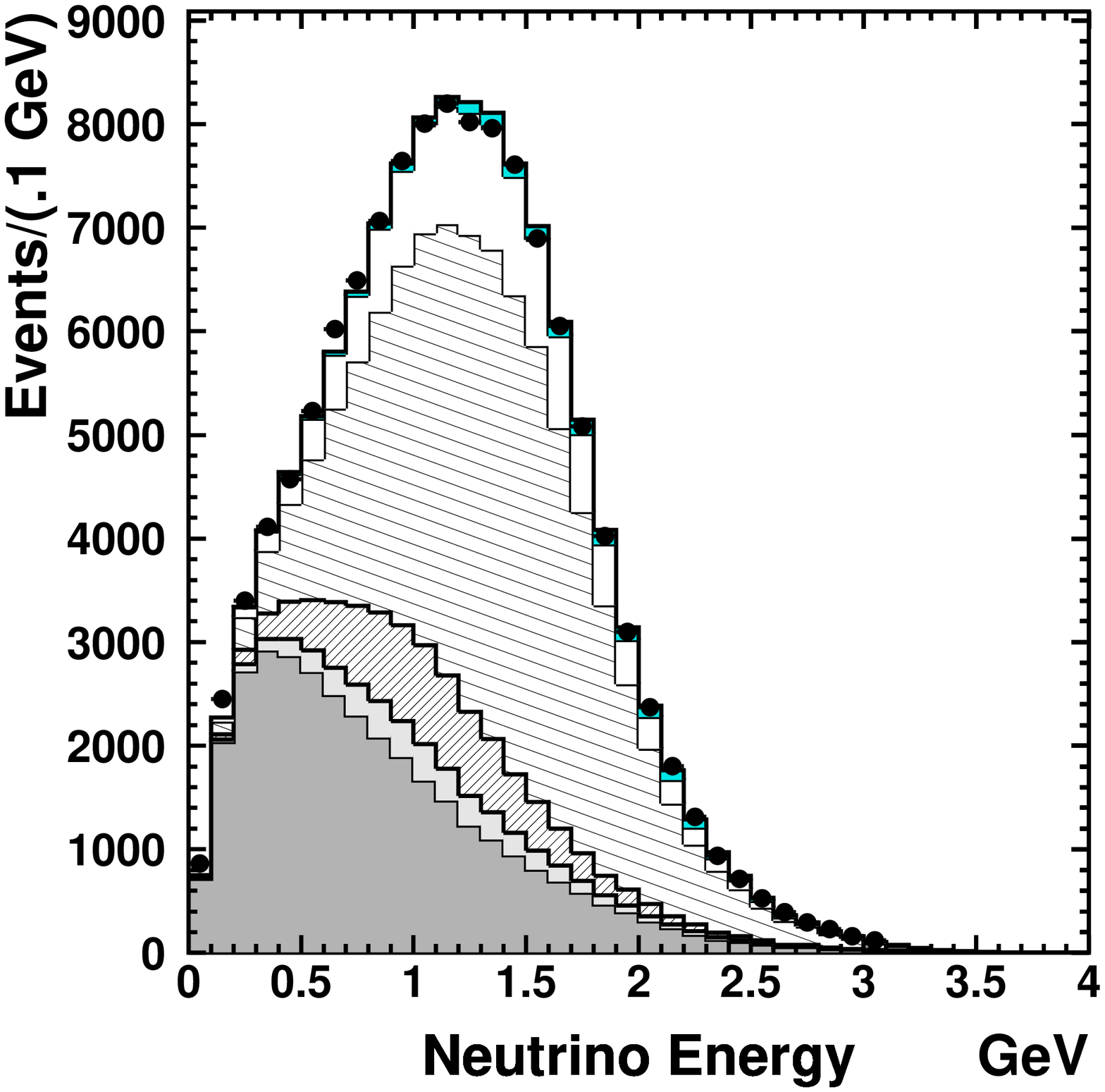}}
  \resizebox{.38\textwidth}{!}{\includegraphics{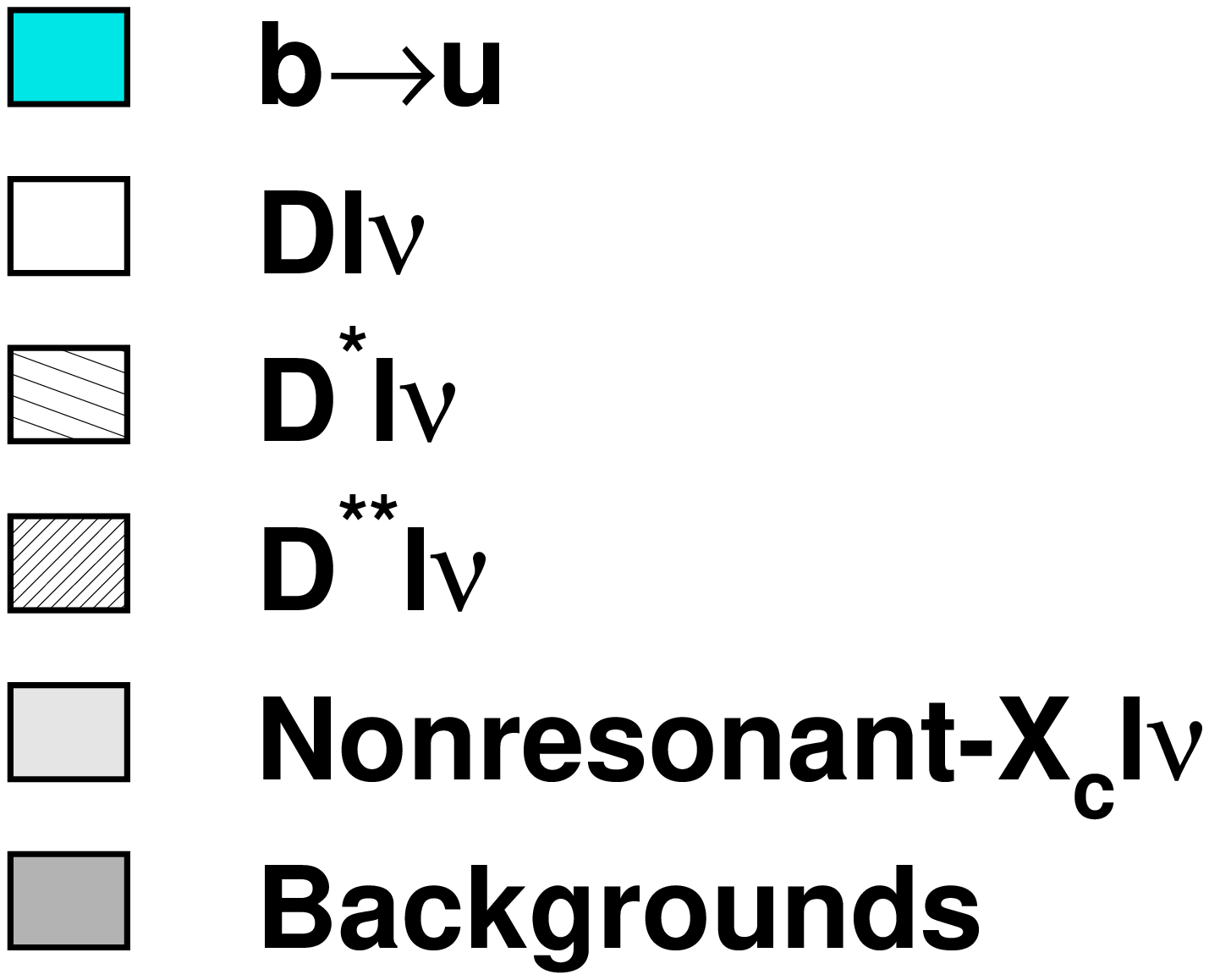}}
}
{
  Projections of the fit results in the reconstructed quantities, on the first row 
  are \QSqr\ and \MXSqr, on the second are \CosWL\ and \Elep, and on the third row is
  \Enu.
  From the top in each figure the histograms are the contributions from \BXulnu\ 
  (barely visible shaded), \BDlnu\ (open), \BDSlnu\ (sparsely hatched),   \BDSSlnu\ (densely hatched), 
  \BXclnu\ nonresonant (light shaded), and the sum of the backgrounds: continuum, secondary leptons,
  fake leptons (dark shaded). The step in the lepton momentum distribution is due 
  to the looser muon identification used below 1.5 \MomUnit\ for which there is a higher fake
  rate and higher efficiency.
}

The \BXlnu\ modes, secondary leptons and real leptons from the continuum are modeled 
with events from a GEANT \cite{ref:GEANT} simulation of the CLEO detector that
are reconstructed in the same manner as data events.
The \BDlnu\ and \BDSlnu\ modes are simulated with HQET using the PDG \cite{ref:PDG}
averages of measurements of the form factors. The \BDSSlnu\ and \BXulnu\ modes
are simulated using form factors from the ISGW2 model \cite{ref:isgw2}. 
The $X_c$ nonresonant modes are simulated with the Goity and Roberts model \cite{ref:GoityRoberts}
in which the $D$ and $D^*$ contributions are excluded. 


The fake leptons are modeled with data events where a track is
selected to be treated as a lepton. The events are then
unfolded bin by bin to extract the $\pi$ and $K$ contributions,
which are then multiplied by the measured fake rates.
This models fake leptons from both \BBbar\ and continuum. This method
also provides an absolute normalization for the fake lepton contribution to the data
sample. The real leptons
from continuum are modeled with a Monte Carlo simulation which has been tuned to 
replicate the appropriate charm spectra; charm is the source of most leptons from 
continuum. The models of both continuum and fakes have been validated and constrained
by comparisons with the $4.5\ {\rm fb}^{-1}$ of off-resonance data. The secondary leptons are modeled 
with CLEO's generic \BBbar\ Monte Carlo which has also been tuned to replicate measured
charm spectra and semileptonic charm decay measurements. To address the issue of the
poorly understood charm counting in $B$ decays, the charm counting in the \BBbar\ Monte Carlo
simulation is tuned to saturate the theoretically predicted level of charm production \cite{ref:PDGold}.

The \BXlnu\ models were varied to assess the dependence of the results on the model.
The \BDlnu\ and \BDSlnu\ modes were varied within the range of the errors on the 
form factor measurements. The curvature of the form factors has not been measured and is
usually constrained by theoretical predictions \cite{ref:FFCurvatures}. Because the
data have an excess above the model in the \QSqr\ region between 5.0 and 8.0 \MassSqrUnit,
the curvature was set to 50\% of its predicted value and varied by 
$\pm$50\% of the prediction. The \BDSSlnu\ form factor was replaced by a model
inspired by HQET calculations \cite{ref:HQETDSSlnu}. The slope of the \BDSSlnu\ and
\BXclnu\ nonresonant form factors in the \QSqr\ dimension was also varied.
The dominant model dependence is from the hadronic mass distribution
of the \BXclnu\ nonresonant mode. This is conservatively reweighted with a series 
of Gaussians restricted to the kinematically allowed region.
The means of the Gaussians are allowed to range from $M_D+M_\pi$ to 3.5 \MassUnit\ with variances
ranging from $0.25$ \MassSqrUnit\ to $1.25$ \MassSqrUnit. The \BXulnu\ simulation is varied 
from an all-nonresonant model to the nominal ISGW2 model \cite{ref:isgw2}, with a hybrid of the two
in between. The all-nonresonant model differential decay rate corresponds to the prediction 
of HQET combined with CLEO's measurement of the \bsg spectral function 
\cite{ref:btoutheory,ref:bsgmeasurement}. The maximum deviation of the \BXclnu\ nonresonant 
mass Gaussians is added in quadrature with the deviation of the other model 
variations to get the total model dependence.

Radiative corrections play an important role in the measurement of the \MXSqr\ 
distribution. The reconstructed \MXSqr\ is defined to be the mass squared of 
the system recoiling against the charged lepton and the neutrino. If a photon
is radiated by the lepton in the event, it will be included in this definition
of the recoil system,
\begin{eqnarray}
   \MXSqr = (p_B^\mu-p_\ell^\mu-p_\nu^\mu)^2 = (p_{h}^\mu+p_\gamma^\mu)^2.
\end{eqnarray}
The goal of this analysis is to measure the mass squared moment of the 
recoiling hadronic system $p_{h}^2$, not its combination with
the radiated photon $(p_{h}+p_\gamma)^2$. To correct for this effect,
the data are fit using fully simulated GEANT Monte Carlo events in which the 
PHOTOS package \cite{ref:PHOTOS} has been used to generate radiated photons. 
The moments are calculated from the fit results and the theoretical models
of the hadronic mass distributions for each mode and thus do not have a shift
due to the radiative corrections.

However, knowledge of the radiative corrections is not complete. PHOTOS implements
an algorithm based on a splitting function which applies the same physics at
$\cal O(\alpha)$ as the prescription of Atwood and Marciano \cite{ref:atwood_marciano}.
PHOTOS also includes a prescription for forcing the kinematics of the decay to
conserve momentum in addition to energy. The implementation of PHOTOS has been checked in
detail against a private implementation of the same algorithm. The applicability
of the PHOTOS algorithm is however not exact. The calculation ignores the internal
structure of the hadronic system. Richter-W\c{a}s \cite{ref:richterwas} has made a comparison of the PHOTOS 
and Atwood-Marciano prescriptions with an exact order-$\alpha$ calculation of the
radiative corrections to the $B^- \ra D^0 \ell^- \nubar$ differential decay rate 
and has found agreement at the 20\% and 30\% level, respectively. 
Because we must extrapolate to the other \BXclnu\ modes, we make a conservative
estimate that the PHOTOS calculation can be trusted to $\pm$50\%.

The application of the radiative corrections is further complicated by the
fact that the generated photons are low energy and often lost. When the photon
is lost it can cause the event to fail the missing mass cut. If the event
does pass the missing mass cut, the reconstructed neutrino will be biased toward high
energy, pushing the reconstructed \MXSqr\ toward the true hadronic mass squared without
the photon.
If neglected, this would increase the measured \MomMXSqrmMDbSqr\ moment with a 1.0
\GeVUnit\ lepton energy cut by 0.082 \MassSqrUnit, before detector effects are included.
This would be reduced to 0.037 \MassSqrUnit\ after detector effects.

The method of neutrino reconstruction adds a large amount of kinematic information
to each event. However, it also adds significant potential for systematic errors. The
resolution on the neutrino kinematics is affected by the models of the signal,
the other $B$ in the event, and the detector response.  The GEANT Monte Carlo simulation
does not perfectly reproduce the track and
shower efficiencies and fake rates, nor are $B$ decays well enough understood that the
inclusive particle distributions are well known. For this analysis we employ
a reweighting method in order to quantify the effects of these uncertainties on our results.
For example, to study the effect of the tracking efficiency uncertainty, the Monte Carlo 
events in which tracks are lost are given a higher or lower weight in constructing the 
component histograms for the fit. The scale of the variation can in general
be constrained by direct measurements of the quantity being varied. One important example is that 
having a $K^0_L$ in the event adds a tail to the neutrino resolution. The inclusive $K^0_S$
spectrum in \BBbar\ events has been measured and can be used to constrain the $K^0_L$ spectrum. 
The dominant detector systematics are fake showers (generally splitoffs from hadronic showers), 
final state radiation, and fake leptons. The detector systematics are summed in quadrature
to arrive at the overall detector systematic error.

\section{Results}

The fit results can be used to calculate a branching fraction for each of the
hadronic final states \BDlnu, \BDSlnu, \BDSSlnu, and \BXclnu\ nonresonant.
From these branching fractions the moment \MomMXSqrmMDbSqr\ with a certain lepton 
energy cut can be calculated as follows:
\begin{eqnarray}
\label{eqn:mommath}
\MomMXSqrmMDbSqr_{cut} = \frac
        {\sum_{m} m_m c_m \BRsym_m    } 
	{\sum_{m}  c_m \BRsym_m },
\end{eqnarray}
where $m_m$ is the moment of mode $m$ with the cut applied, $c_m$ is the
fraction of mode $m$ passing the cut, and $\BRsym_m$ is the branching fraction
of mode $m$. The quantities $m_m$ and $c_m$ depend only on the model. The measured
branching fractions, $\BRsym_m$, depend on the model, the detector simulation,
and the data.

For each lepton energy cut, the fit is repeated with the cut applied using the 
reconstructed (lab frame) quantity. Theoretical calculations of 
\MomMXSqrmMDbSqr, however, are made with 
the cut in the $B$ rest frame \cite{ref:falkcut,ref:bauercut}. The moments
reported are therefore calculated in the $B$ rest frame. This involves a small
extrapolation. 

The resulting moments are shown in Table \ref{tab:momvscut}. These
moments are highly correlated because they share some fraction of
the data and also use the same models and detector simulation. The
statistical correlations are assessed with a simple Monte Carlo simulation.
The model dependence and detector systematic correlations are assessed
by summing the covariance matrices from each systematic error. The resulting
covariance matrix is shown in Table \ref{tab:covariance}. An alternative
representation of the data is to use the \MomMXSqrmMDbSqr\ moment with a 
1.5 \GeVUnit\ lepton energy cut and the difference between the \MomMXSqrmMDbSqr\ moments
at 1.0 and 1.5 GeV. These two measurements are significantly less correlated.
We find 
\[
\MomOf{\MXSqr}_{\Elep>1.0\,{\rm GeV}}-\MomOf{\MXSqr}_{\Elep>1.5\,{\rm GeV}} =
(0.163 \pm 0.014 \pm 0.036 \pm 0.064)\,\MassSqrUnit
\] 
and a covariance of this value with the 
$\MomMXSqrmMDbSqr_{\Elep>1.5\,{\rm GeV}}$ moment of $2.242\times10^{-3}\,\MassFourthUnit$.

\begin{table}[hptb]
\caption{\MomMXSqrmMDbSqr\ versus the lepton energy cut. The errors on the entries
in the table are the statistical, detector systematics, and model dependence, respectively.}
\label{tab:momvscut}
\smallskip
\begin{tabular}{c|c}
Cut (\GeVUnit)& \MomMXSqrmMDbSqr~(\MassSqrUnit)\\
\hline
\hline
$\Elep > 1.0 $~  &~ 0.456 $\pm$ 0.014 $\pm$ 0.045 $\pm$ 0.109 \\ 
$\Elep > 1.1 $~  &~ 0.422 $\pm$ 0.014 $\pm$ 0.031 $\pm$ 0.084 \\ 
$\Elep > 1.2 $~  &~ 0.393 $\pm$ 0.013 $\pm$ 0.027 $\pm$ 0.069 \\ 
$\Elep > 1.3 $~  &~ 0.364 $\pm$ 0.013 $\pm$ 0.030 $\pm$ 0.054 \\ 
$\Elep > 1.4 $~  &~ 0.332 $\pm$ 0.012 $\pm$ 0.027 $\pm$ 0.055 \\ 
$\Elep > 1.5 $~  &~ 0.293 $\pm$ 0.012 $\pm$ 0.033 $\pm$ 0.048 \\ 
\end{tabular}
\end{table}

\begin{table}[hptb]
\caption{Covariance of \MomMXSqrmMDbSqr\ moments with different lepton energy cut.
The rows/columns correspond to progressively more restrictive lepton energy cut starting
with 1.0 \GeVUnit\ and ending with 1.5 \GeVUnit. }
\label{tab:covariance}
\[
\left(
\begin{matrix}
1.421  &  1.096  &  0.913  &  0.724  &  0.735  &  0.618  \cr
1.096  &  0.864  &  0.719  &  0.577  &  0.574  &  0.486  \cr
0.913  &  0.719  &  0.612  &  0.495  &  0.499  &  0.422  \cr
0.724  &  0.577  &  0.495  &  0.413  &  0.418  &  0.352  \cr
0.735  &  0.574  &  0.499  &  0.418  &  0.447  &  0.373  \cr
0.618  &  0.486  &  0.422  &  0.352  &  0.373  &  0.324  \cr
\end{matrix} 
\right) \times 10^{-2}~\MassFourthUnit
\]
\end{table}

Figure \ref{fig:momvsth} shows a comparison of the results with previous
measurements \cite{ref:oldhadmom,ref:babarmoms} and a prediction based on HQET 
constrained by this measurement
of $\MomMXSqrmMDbSqr_{\Elep>1.5}$ and the first \bsg photon energy moment 
\cite{ref:bsgmeasurement}. The results agree well with both.
The theory bands shown in the figure reflect the 
experimental errors on the two constraints, the variation of the third-order 
HQET parameters by the scale $(0.5 \ \GeVUnit)^3$, and variation of the size of the 
higher order QCD radiative corrections \cite{ref:bauercut}.

\mypsf{fig:momvsth}
{
\resizebox{.6\textwidth}{!}{\includegraphics{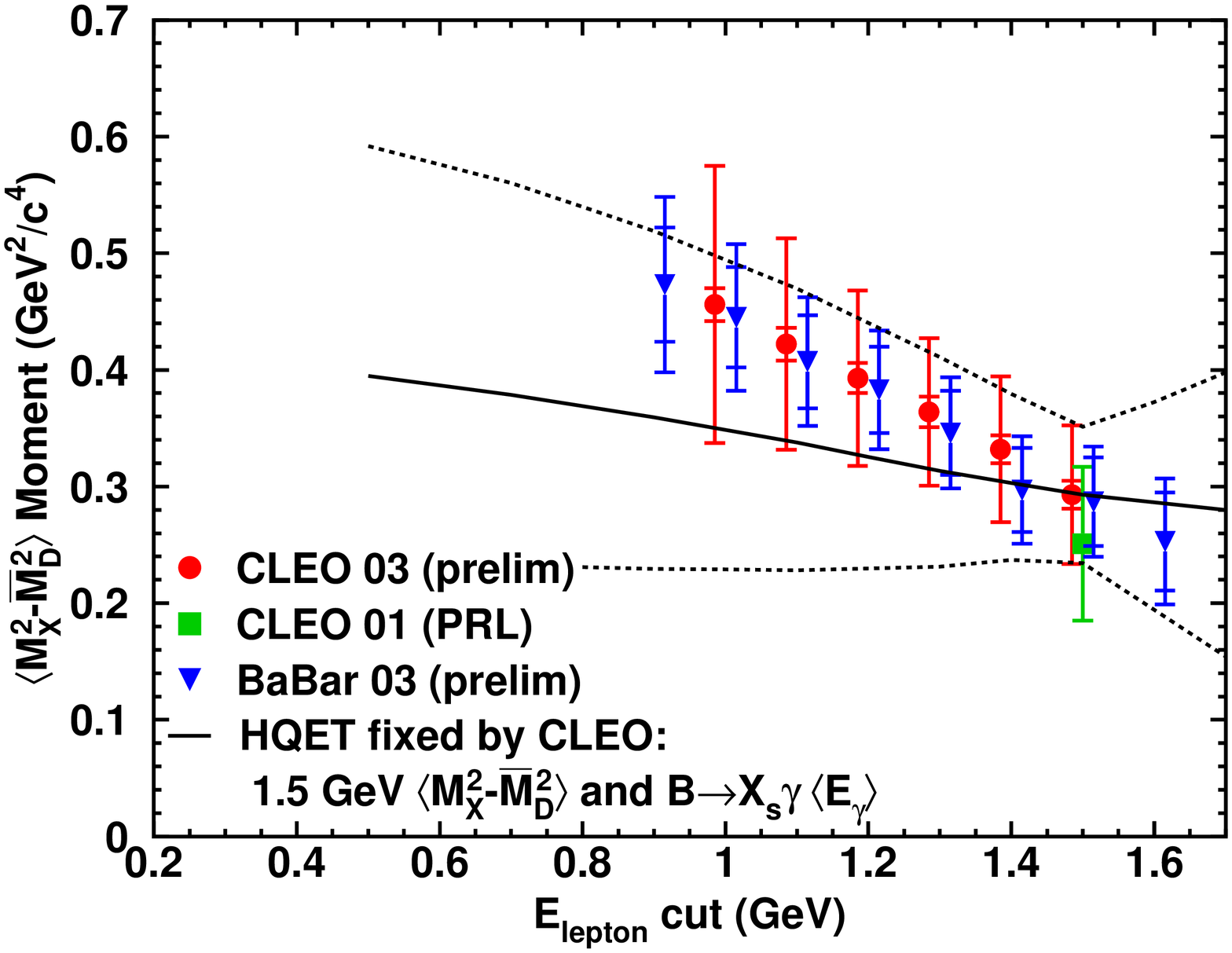}}
}
{
The results of this analysis compared to previous measurements 
\cite{ref:oldhadmom,ref:babarmoms} and the HQET prediction. The theory bands shown in 
the figure reflect the variation of the experimental errors
on the two constraints, the variation of the third-order HQET parameters by
the scale $(0.5\ \GeVUnit)^3$, and variation of the size of the higher order QCD
radiative corrections \cite{ref:bauercut}.
}

\section{Conclusion}

We have measured the moment \MomMXSqrmMDbSqr\ as a function of the 
lepton energy cut with a minimum cut of 1.0 \GeVUnit. We obtain the preliminary result,
\begin{eqnarray*}
\MomMXSqrmMDbSqr_{\Elep>1.0\,{\rm GeV}} =  (0.456 \pm 0.014 \pm 0.045 \pm 0.109)\ \MassSqrUnit.
\end{eqnarray*}
\noindent
This result provides a test of the HQET predictions using a larger portion
of the available decay rate. We have also improved the precision of the 
measurement with a 1.5 \GeVUnit\ lepton energy cut, obtaining the preliminary result
\begin{eqnarray*}\MomMXSqrmMDbSqr_{\Elep>1.5\,{\rm GeV}} =  (0.293 \pm 0.012 \pm 0.033 \pm 0.048)\ \MassSqrUnit.
\end{eqnarray*}
\noindent
Measurements such as these can be used to improve the determination of the 
Standard Model parameter \Vcb. They also provide a 
valuable test of the HQET inclusive predictions which are used in the
extraction of \Vub\ and \Vcb\ and in the interpretation \bsg.

We gratefully acknowledge the contributions of the CESR staff for providing the luminosity
and the National Science Foundation and U.S. Department of Energy for supporting this research.


\begin{thebibliography}{99}

\bibitem{ref:general_HQET_OPE}
J.~Chay, H.~Georgi and B.~Grinstein,
Phys.\ Lett.\ B {\bf 247}, 399 (1990);
I.~I.~Bigi, M.~A.~Shifman, N.~G.~Uraltsev and A.~I.~Vainshtein,
Phys.\ Rev.\ Lett.\  {\bf 71}, 496 (1993)
[arXiv:hep-ph/9304225];
 A.~V.~Manohar and M.~B.~Wise,
Phys.\ Rev.\ D {\bf 49}, 1310 (1994).

\bibitem{ref:oldhadmom}
S.~E.~Roberts, PhD thesis, University of Rochester, 
UMI-98-08907;
D.~Cronin-Hennessy {\it et al.}  [CLEO Collaboration],
Phys.\ Rev.\ Lett.\  {\bf 87}, 251808 (2001)
[arXiv:hep-ex/0108033].

\bibitem{ref:babarmoms}
B.~Aubert, {\it et~al.} [BaBar Collaboration], hep-ex/0307046.

\bibitem{ref:firstnurec}
J.~P.~Alexander {\it et al.}  [CLEO Collaboration],
Phys.\ Rev.\ Lett.\  {\bf 77}, 5000 (1996).

\bibitem{ref:falkcut}
A.~F.~Falk and M.~E.~Luke,
Phys.\ Rev.\ D {\bf 57}, 424 (1998)
[arXiv:hep-ph/9708327].

\bibitem{ref:NIM}
Y.~Kubota {\it et~al.}, {Nucl. Instrum. Meth. A} \textbf{320}, {66} ({1992});
T.~S.~Hill, {Nucl. Instrum. Meth. A} \textbf{{418}}, {32} ({1998}).

\bibitem{ref:GEANT}
R.~Brun {\it et~al.}, GEANT 3.15, CERN Report No. DD/EE/84-1 (1987).

\bibitem{ref:foxwolfram}
G.~C.~Fox and S.~Wolfram,
Phys.\ Rev.\ Lett.\  {\bf 41}, 1581 (1978).

\bibitem{ref:BarlowBeeston}
R.~J.~Barlow and C.~Beeston,
Comput.\ Phys.\ Commun.\  {\bf 77}, 219 (1993).


\bibitem{ref:PDG}
K.~Hagiwara {\it et al.}  [Particle Data Group Collaboration],
Phys.\ Rev.\ D {\bf 66}, 010001 (2002).

\bibitem{ref:isgw2}
D.~Scora and N.~Isgur,
Phys.\ Rev.\ D {\bf 52}, 2783 (1995)
[arXiv:hep-ph/9503486].

\bibitem{ref:GoityRoberts}
J.~L.~Goity and W.~Roberts,
Phys.\ Rev.\ D {\bf 51}, 3459 (1995)
[arXiv:hep-ph/9406236].

\bibitem{ref:PDGold}
C.~Caso {\it et al.}  [Particle Data Group Collaboration],
Eur.\ Phys.\ J.\ C {\bf 3}, 1 (1998).

\bibitem{ref:FFCurvatures}
C.~G.~Boyd, B.~Grinstein and R.~F.~Lebed,
Phys.\ Lett.\ B {\bf 353}, 306 (1995)
[arXiv:hep-ph/9504235];
I.~Caprini and M.~Neubert,
Phys.\ Lett.\ B {\bf 380}, 376 (1996)
[arXiv:hep-ph/9603414];
C.~G.~Boyd, B.~Grinstein and R.~F.~Lebed,
Phys.\ Rev.\ D {\bf 56}, 6895 (1997)
[arXiv:hep-ph/9705252];
I.~Caprini, L.~Lellouch and M.~Neubert,
Nucl.\ Phys.\ B {\bf 530}, 153 (1998)
[arXiv:hep-ph/9712417].

\bibitem{ref:HQETDSSlnu}
A.~K.~Leibovich, Z.~Ligeti, I.~W.~Stewart and M.~B.~Wise,
Phys.\ Rev.\ D {\bf 57}, 308 (1998)
[arXiv:hep-ph/9705467].

\bibitem{ref:btoutheory}
M.~Neubert,
Phys.\ Rev.\ D {\bf 49}, 3392 (1994)
[arXiv:hep-ph/9311325];
M.~Neubert,
Phys.\ Rev.\ D {\bf 49}, 4623 (1994)
[arXiv:hep-ph/9312311];
I.~I.~Bigi, M.~A.~Shifman, N.~Uraltsev and A.~I.~Vainshtein,
Phys.\ Lett.\ B {\bf 328}, 431 (1994)
[arXiv:hep-ph/9402225];
A.~L.~Kagan and M.~Neubert,
Eur.\ Phys.\ J.\ C {\bf 7}, 5 (1999)
[arXiv:hep-ph/9805303];
F.~De Fazio and M.~Neubert,
JHEP {\bf 9906}, 017 (1999)
[arXiv:hep-ph/9905351];
A.~K.~Leibovich, I.~Low and I.~Z.~Rothstein,
Phys.\ Rev.\ D {\bf 61}, 053006 (2000)
[arXiv:hep-ph/9909404].


\bibitem{ref:bsgmeasurement}
S.~Chen {\it et al.}  [CLEO Collaboration],
Phys.\ Rev.\ Lett.\  {\bf 87}, 251807 (2001)
[arXiv:hep-ex/0108032].


\bibitem{ref:PHOTOS}
E.~Barberio and Z.~W\c{a}s,
Comput.\ Phys.\ Commun.\  {\bf 79}, 291 (1994).

\bibitem{ref:atwood_marciano}
D.~Atwood and W.~J.~Marciano,
Phys.\ Rev.\ D {\bf 41}, 1736 (1990).

\bibitem{ref:richterwas}
E.~Richter-W\c{a}s,
Phys.\ Lett.\ B {\bf 303}, 163 (1993).

\bibitem{ref:bauercut}
C.~W.~Bauer, Z.~Ligeti, M.~Luke and A.~V.~Manohar,
Phys.\ Rev.\ D {\bf 67}, 054012 (2003)
[arXiv:hep-ph/0210027].


\end{thebibliography}
\end{document}